\documentclass[12pt]{article}
\usepackage{amsmath}
\numberwithin{equation}{section}
\usepackage[left=0.95in,right=0.95in 
]{geometry}
\usepackage{setspace}
\usepackage [linktocpage]{hyperref}
\date{(First dated: March 20, 2011)}
\begin{document}
\title{\bf A Monopole Instanton-Like Effect in the ABJM Model \ \\ \ }
 \author{M. Naghdi\footnote{m.naghdi@kurdland.com} \\ 
\textit{Department of Physics, School of Sciences}, \\
\textit{Tarbiat Modares University, Tehran, Iran} } 
 \setlength{\topmargin}{0.2in}
 \setlength{\textheight}{9.2in}
  \maketitle
   \vspace{0.3in}
    \thispagestyle{empty}
   \begin{center}
   \textbf{Abstract}
   \end{center}

   Making use of ansatzs for the form fields in the 10d type IIA supergravity version of the ABJM model, we come with a solution in the Euclidean signature recognized as a monopole instanton-like object. Indeed we will see that we can have a (anti) self-dual solution at a special limit. While as a topological object, its back-reaction on the original background should be ignorable, we show the energy-momentum tensors vanish exactly. On the field theory side, the best counterpart is an U(1) gauge field of a gauge transformation. To adjust with bulk, the gauge field must prompt to a dynamic one without adding any kinetic term for this dual photon except a marginal, abelian AB-type Chern-Simons term on the boundary. We will see how both side solutions match next to another confirmation from some earlier works of this vortex-particle duality.

  \newpage
  \setlength{\topmargin}{-0.7in}
  \pagenumbering{arabic} 
  \setcounter{page}{2} 

\section{Introduction}
Nonperturbative effects, known to be Solitons and Instantons mainly, are of particular attentions in the gauge and string theories. Even more intersecting ones are instantons, meaning the finite action Euclidean solutions to the equations of motion. Searching for their existence in the ABJM model \cite{ABJM}, as the best so far known sample of AdS$_4$/CFT$_3$ duality, has seriously been started in \cite{Hosomichi}. They studied a special case of the Euclideanized (M) D-branes. We have also studied \cite{I.N} another particular case of M-branes recently. In the following, we present another sample of D-instantons.

The ABJM theory \cite{ABJM} states the near horizon limit of NM2-branes, which probe a $C^4/Z_k$ singularity in M-theory, is dual to a three-dimensional conformal $\mathcal{N}=6$ Chern-Simon-matter $G=U(N)_k \times U(N)_{-k}$ Yang-Mills field theory living on the boundary of $AdS_4$. There are $k$ as the Chern level, matters in bi-fundamental representations of the gauge group and $N$ unit of 4-form flux in the bulk of $AdS_4$. On the other hand, by breaking up the isometry group as $SO(8) \rightarrow SU(4) \times U(1)$, one may consider $S^7$ as a $U(1)$ fiber bundle over $CP^3$. When $k$ increases the M-theory circle becomes small and a good description is by type IIA theory on $AdS_4 \times CP^3$. Then, there are $N$ unit of $F_4$ flux on $AdS_4$, $k$ unit of $F_2$ flux on 2-cycle $CP^1 \subset CP^3$ and $H_3=dB_2=0$, where $B_2$ is NSNS 2-form field. Indeed, this supergravity approximation is valid when the 't Hooft coupling $\lambda \equiv N/k \gg 1$  and $k^5 \gg N$.

We try to add some terms- indeed forms- to the original field-strengths while preserving other ABJM backgrounds such as metric, dilaton and other fields as original. In general, these ansatzs can be guessed according to the standard brane-solution building manner. There always is a scalar field playing role as fluctuations on the branes and often identifies with the localized objects in the bulk of $AdS_4$. Dual dynamics, on the boundary, may realize with the scalars, gauge fields or fermions.

In fact, here, we meet a massless U(1) gauge field in the bulk of AdS and not a scalar, which is the usually faced case. For a massless U(1) gauge field in the bulk of $AdS_4$, dual boundary operators have the conformal dimensions $\Delta_\pm=1,2$ similar to the case of a scalar with the mass squared $m^2=-2$. We have already been considering an instance of the latter in \cite{I.N}, which is indeed a conformally coupled scalar. With gauge fields, both modes are renormalizable contrary to the scalar case where just the upper branch is renormalizable. On the other hand, depended on the boundary condition, an AdS theory with a U(1) gauge field in 4d has an infinite number of dual boundary CFT's. Similar to scalars in AdS, there are "magnetic" and "electric" boundary conditions for gauge fields equivalent to the Dirichlet and Neumann boundary conditions for scalars, respectively. A fixed magnetic field on the boundary equals the Dirichlet condition stating the gauge field vanishes on the boundary up to some gauge transformations. In the latter case, the magnetic charge is forbidden in the bulk but the net electric charge is equivalent to a conserved quantity on the boundary. Opposite to this is right for the Neumann boundary condition. $\vec{E}=0$ theory stands from $\vec{B}=0$ by coupling of a gauge field $A$, without any kinetic energy, to a conserved current $J$. In general, as first discussed in \cite{Witten2} and further studied in \cite{Yee}, after an arbitrary $SL(2,Z)$ transformation of the boundary conditions, a combination of electric and magnetic charges in the bulk is allowed replying to the conserved charge of the boundary theory. The S-operation of $SL(2,Z)$, corresponding to this electric-magnetic duality, is nothing but the Legendre transformation when going from one CFT having the operator $\Delta_+$ (Dirichlet  boundary condition) to another one having the operator $\Delta_-$ (Neumann boundary conditions) as surveyed further in \cite{deHaro}. We will see another exact example of this S-duality at follows.

Rest of this note organizes as follows. In section 2, we discuss on the gravity side. In subsection 2.1, we review the needed subjects of the 10d type IIA supergravity version of the ABJM model and some general statements about solutions and equations of motion to be satisfied in general. In subsection 2.2, we present our ansatz with some gravity side of the solution. We show, for a special case of (anti) self-dual solution, there is not any back-reaction on the original background because of the added effect. Next, we calculate the correction impelled by this Euclideanized object next to its charge. Section 3 assigns to the field theory side. There we review the needed material of the field theory of ABJM. Then we see how matching with the gravity side solution by someways, such as symmetries, hint us to find the wished dual boundary solution and operator. In Section 4 are some closing remarks stressing on the electric-magnetic duality of our solution.

\section{Gravity Side Solution}
\subsection{General Remarks}
The 10-dimensional type IIA supergravity action in string frame is given by
\begin{equation}\label{eq1}
\begin{split}
  S_{IIA} = & \frac{1}{2 \kappa^2} \int d^{10}x \, \sqrt{g} \, e^{-2\phi} \, R+\frac{1}{2 \kappa^2} \int \biggl\lbrack e^{-2\phi}  \, \bigl(4 d\phi \wedge \ast d\phi-\frac{1}{2} H_3 \wedge \ast H_3 \bigr) \\
  & -\frac{1}{2} F_2 \wedge \ast F_2-\frac{1}{2} \widetilde{F}_4 \wedge \ast \widetilde{F}_4 -\frac{1}{2} B_2 \, \wedge F_4  \, \wedge F_4 \biggr\rbrack
\end{split}
\end{equation}
where $H_3=dB_2,\ F_2=dA_1,\ F_4=dA_3,\ \widetilde{F}_4=dA_3-A_1 \wedge H_3$ and the Hodge-star operation is with respect to full 10d metric. The corresponding ABJM \cite{ABJM} geometry in string frame (in unit where $\acute{\alpha}=1$) is
\begin{equation}\label{eq2}
   ds_{ABJM(IIA)}^2 = \tilde{R}^2 \big(ds_{AdS_4}^2+4ds_{CP^3}^2 \big), \quad \tilde{R}^2 =\frac{R^3}{4k}=\pi \sqrt{\frac{2N}{k}}=\pi \sqrt{2\lambda}
\end{equation}
in which $ds_{AdS_4}^2$, $ds_{CP^3}^2$ are unit-radius metrics of the associated spaces, $\lambda\equiv N/k$ is 't Hooft coupling and $R=2L$ is the AdS curvature radius in the Poincare upper-half plane coordinate we use here. In the latter frame with Euclidean signature, we have for $AdS_4$
\begin{equation}\label{eq3}
ds^2_{EAdS_4}=\frac{L^2}{u^2} \big(du^2+ dx_i dx_i \big), \quad i=1,2,3 
\end{equation}
In fact, type IIA supergravity approximation on $AdS_4 \times CP^3$ is valid when $\lambda \equiv N/k \gg 1$  and $k^5 \gg N$. Dilation and field-strengths (forms)-with $N$ units of 6-form flux on $CP^3$ and $k$ units of 2-form flux on $CP^1\subset CP^3$ read as well
\begin{equation}\label{eq4}
  e^{2\phi}=\frac{R^3}{k^3}, \quad H_3=0, \quad F_2^{(0)}=dA_1^{(0)}=kJ, \quad F_4^{(0)}=dA_3^{(0)}=\frac{3}{8} R^3 \mathcal{E}_{AdS_4}\equiv \tilde{N} \mathcal{E}_4
\end{equation}
where $\mathcal{E}_4$ is the $AdS_4$ unit volume-form and $J$ is proportional to the K\"{a}hler form on $CP^3$.

By taking $H_3=0$ as in the ABJM, the formic relations to satisfy are
\begin{equation}\label{eq5a}
   dF_{p}=0,\quad d\ast F_{p}=0,
\end{equation}
\begin{equation}\label{eq5b}
   d\ast H_3 = g_s^2 (-F_2\wedge \ast \widetilde{F}_4+\frac{1}{2} \widetilde{F}_4 \wedge \widetilde{F}_4)=0
\end{equation}
where $p=2,4$ and that in (\ref{eq5b}) the use is made of the fact that, as in the ABJM, dilaton is constant with $e^{2\phi}=g_s^2$. Next to above, satisfying the dilaton and Einstein equations are required. In fact, the dilaton equation
\begin{equation}\label{eq6}
    d(\ast d\phi)-d\phi \wedge \ast d\phi-\frac{1}{8}H_3 \wedge \ast H_3+\frac{1}{4.3!} R \mathcal{E}_4 \wedge J^3 =0
\end{equation}
is satisfied automatically because we don't change the original background except $F_p$'s. Therefore, the dilaton $\phi$ is still constant, $H_3=0$ and the Ricci scalar $R_{AdS_4\times CP^3}$ vanishes for the ABJM geometry as well. The only remained relation to satisfy is the RHS of the Einstein equation - on which are the energy-momentum tensors
\begin{equation}\label{eq7}
    R_{MN}-\frac{1}{2} g_{MN} R=-8T_{MN}^\phi+ T_{MN}^{H_3}+e^{+2\phi} T_{MN}^{F_2}+e^{+2\phi} T_{MN}^{\widetilde{F}_4}
\end{equation}
where the capital indices $M, N$ here are for the 10d space-time directions. As long as we search for topological objects, it is pleasure the energy-momentum tensors of the added fields vanish.  This guarantees new effects don't back-react on the geometry. Although this settles for our solution but it doesn't in general happen. As a simple way to resolve this, we first note that $k$ becomes large in reducing to the type IIA version (indeed $k^5 \gg N$) and that $e^{2\phi}=g_s^2=R^3/k^3$ in ABJM. Then, we are seemingly able to ignore all $F_p$ terms on the RHS of the Einstein equation in string frame as both terms in (\ref{eq7}) have an $e^{2\phi}$ prefactor and same for the RHS of $B_2$ equation up to some approximation. A more standard manner is that surveyed in \cite{Skenderis} for instance. According to that, as long as we are interested in the behavior of solutions near to the boundary and gravitational and field equations decouple at $u\to 0$, we can safely ignore back-reactions and study each field in a fixed background. This is case for our solution as well. Nevertheless, still another common way is simply making use of "probe approximation" that is neglecting the back-reactions of the added objects on the original geometry. The argument for the latter is since the background is such strong that adding few weak effects doesn't change it drastically. This confirms for our ansatz of course. Therefore, altogether, our ansatzs here just need to obey (\ref{eq5a}).

\subsection{D0-D2 branes: Ansatz for $F_2-F_4$}
Here we consider ansatzs for $F_2$ from two related manners leading to a same result. The configuration is a massless U(1) gauge field in the bulk whose excitation induces a magnetic field on the boundary of $AdS_4$. The final solution may be interpreted as a monopole in the Euclidian signature we call it a monopole instanton-like effect.

Following the discussion in the ABJM, for the U(1) massless states in the bulk, we make the following ansatz  
\begin{equation}\label{eq8}
   F_2=kJ+\grave{k}F^{D0}, \qquad {F}_4=-i\tilde{N}\mathcal{E}_4 \mp i J \wedge \widetilde{F}^{D2}, \qquad H_3=0
\end{equation}
where $F^{D0}$ and $\widetilde{F}^{D2}$ are completely in the $AdS_4$ directions\footnote{A similar ansatz has already been considered about Fractional Quantum Hall Effect (FQHE) in \cite{Hikida}.} and that $\grave{k}$ is a constant whose significance will be clear soon. The $i$ factor is required for working in the Euclidean signature and the two signs $\mp$ are for (anti) self-dual configuration respectively as we will see. Having this ansatz, the relations (\ref{eq5a}) reduce to the following 4d ones
\begin{equation}\label{eq9}
 d\widetilde{F}^{D2}=0, \qquad d\ast_4 \widetilde{F}^{D2}=0, \qquad d\ast_4F^{D0}=0, \qquad dF^{D0}=0
\end{equation}
while the Euclideanized version of the RHS of $B_2$ equation (\ref{eq5b}) reads
\begin{equation}\label{eq10}
   \begin{split}
   -F_2\wedge \ast {F}_4 -i\frac{1}{2} {F}_4 \wedge {F}_4 &= \frac{R^3}{8} \big(+\frac{\grave{k}R^3}{k}F^{D0}\pm 2\ast_4 \widetilde{F}^{D2}\big)\wedge J^3 \\
   & + \big(\pm \frac{\grave{k}R^3}{4k} F^{D0}\wedge \ast_4 \widetilde{F}^{D2}+\frac{1}{2}\widetilde{F}^{D2} \wedge \widetilde{F}^{D2}\big) \wedge J^2=0
 \end{split}
\end{equation}
where we have used the Hodge-duals for $F_2$ and $F_4$ in (\ref{eq8}), with the metric in (\ref{eq2}), as
\begin{equation}\label{eq10b}
  \ast F_2=\frac{R^9}{16.4k^2} J^2 \wedge \mathcal{E}_4+\bigg(\frac{\grave{k}R^9}{48k^3}\bigg) \ast_4 F^{D0} \wedge J^3
\end{equation}
\begin{equation}\label{eq10c}
  \ast F_4=-i\frac{R^6}{8k} J^3 \mp i\frac{R^3}{4k}\ast_4 \widetilde{F}^{D2} \wedge J^2
\end{equation}
and some basic formula for the unit-volumes as \footnote{It is notable that to adjust with our notation, of $ F_2^{(0)}$ in (\ref{eq4}), we take the unit-volume element for $CP^3$ as in (\ref{eq10a}). Then, we should also take $\int_{CP^1}J=2\pi$.}
\begin{equation}\label{eq10a}
  \begin{split}
    & \ \texttt{d}Vol(CP^3)=\frac{1}{3!} \frac{J}{2} \wedge \frac{J}{2} \wedge \frac{J}{2} \equiv \frac{1}{8.3!} J^3, \\
     & \ \ \ \ \ \ \ \ast_6 J= \frac{1}{2.2!} J^2, \qquad J \wedge J^3=0, \\
     & \mathcal{E}_4\wedge \mathcal{E}_4=0, \qquad \mathcal{E}_4 \wedge F^{D0}=\mathcal{E}_4 \wedge \widetilde{F}^{D2}=0
 \end{split}
\end{equation}
The two terms in (\ref{eq10}) must separately vanish. The first one, from left, imposes
\begin{equation}\label{eq11}
 \widetilde{F}^{D2}=\mp (8\pi^2 {\grave{k}}^2 \lambda)^{1/2} \ast_4F^{D0}
\end{equation}
and plug this into the second term implies
\begin{equation}\label{eq12}
 F^{D0} \wedge F^{D0}=0
\end{equation}
To satisfy the last equation, we can take $F^{D0}$ to have indices only along three directions $i,j,\ldots=1,2,3$ of $AdS_4$. For example we may set
\begin{equation}\label{eq13}
 F^{D0}=dA^{D0}=\frac{1}{2} F_{ij}^{D0} dx^i \wedge dx^j
\end{equation}
To solve the equations of motion (\ref{eq9}), we set
\begin{equation}\label{eq14}
 F_{ij}^{D0}=\varepsilon_{ijk} \partial^k f
\end{equation}
with a scalar function $f$ independent of $u$. Note that $F_{i4}^{D0} = 0$, so that the equations of motion are automatically satisfied. However, the Bianchi identity requires
\begin{equation}\label{eq15}
 \partial_i \partial^i f=0
\end{equation}
with the solution
\begin{equation}\label{eq15a}
 f(r)=c_1+\frac{c_2}{r}
\end{equation}
where $c_1$ is the value of $f$ at infinity, $c_1=f_\infty$, $c_2\equiv c$ is a constant proportional to the object charge and $r=\sqrt{|\vec{u}|^2}=\sqrt{x_i x^i}$. This spherically symmetric solution is singular at the origin $r=0$, where it solves with a delta-function source\footnote{D-branes, as solutions to the supergravity equations of motion, have in general delta-function sources. These singularities may resolve in the full string theory.}. Because of this, we may add the source term $\delta^{(3)} (x^i) f^{-1}$ to the action which cancels the singularity in the field equation. This source term tells us that we have a new object in the theory at $r = 0$. We may also call this object a monopole like D0-instanton. Therefore we left with a string of the monopoles laying along the $u$ direction of $AdS_4$. On the other hand, from equation (\ref{eq11}) we have
\begin{equation}\label{eq16}
 \widetilde{F}^{D2}=\mp \frac{\grave{k}R^3}{4k}\partial_i f dx^i \wedge du
\end{equation}

There is another similar way to appear this solution. In fact, one can choose a similar structure to the latter solution as an ansatz first and then check whether it satisfies the required relations or not. In fact, we may write\footnote{A similar example is discussed in \cite{Witten}, where the tangent direction is $x_i$ and not $u$. Indeed if we take $u\rightarrow x_i$ in ansatz (\ref{eq17}), then the equation and solution are again (\ref{eq15}) and (\ref{eq15a}) respectively except for the mentioned interchange of coordinates. We now have a string of monopoles along the axis $x_i$ while the scalar fluctuations are along other directions. For instance, if we take $x_i=z$, without any prefer, then we have monopoles on the boundary $xy$ planes orthogonal to the axis $z$. Further, for the latter case, by introducing the polar coordinates ($r,\theta,\phi$), the straightforward view is taking $\theta=\pi/2$ from which the famous quantization condition of the magnetic charge extracts.}
\begin{equation}\label{eq17}
 \tilde{F}_2=d\bar{f} \wedge du
\end{equation}
satisfying the equation in (\ref{eq5a}), again gives the 3d Laplace equation (\ref{eq15}) on the boundary. From the ansatz structure, obviously it has an interpretation in the context of monopoles. The world-line of this point-like object is along the direction $u$. With the world-volume directions of branes in the type IIA near horizon limit of the ABJM-shown in Table \ref{table1}- and brane intersection rules, one can easily check that is a half-BPS configuration\footnote{It is notable that $D0^b$-brane, in the table, may form a non-threshold BPS bound-state with the original ABJM branes. In other words, for each pair of D-branes, if there is 0, 4 or 8 relative transverse direction, they refer as threshold BPS bound states satisfying the no force condition. While with 2 or 6 for the number of relative transverse direction of branes, they refer as non-threshold BPS bound states as hinted in \cite{9612095}.}.
\begin{table}[htp]
\centering
\begin{tabular}{||c||c c c |c| c c c c c c|c||}
\hline
\textbf{$AdS_4 \times CP^3$} & \textbf{$x$} & \textbf{$y$} & \textbf{$z$} & \textbf{$u$} & \textbf{$\theta_1$} & \textbf{$\varphi_1$} & \textbf{$\theta_2$} & \textbf{$\varphi_2$} & \textbf{$\xi$} & \textbf{$\psi$} & $\neq$\\
\hline
\hline
\textbf{ND2} & \textbf{--} &\textbf{--} & \textbf{--} & \textbf{$\times$} & \textbf{$\times$} & \textbf{$\times$} & \textbf{$\times$} & \textbf{$\times$} & \textbf{$\times$} & \textbf{$\times$} & \textit{$n_1$} \\
\textbf{kD6-flux} & \textbf{--} & \textbf{--} & \textbf{--} & \textbf{$\times$} & \textbf{$\times$} & \textbf{$\times$} & \textbf{--} & \textbf{--} & \textbf{--} & \textbf{--} & \textit{$n_2$}\\
\hline
\hline
\textbf{$D0^a$} & $\times$ & $\times$ & $\times$ & -- & $\times$ & $\times$ & $\times$ & $\times$ & $\times$ & $\times$ & \textit{$n_1$}:4, \textit{$n_2$}:8\\
\textbf{$D0^b$} & $\times$ & $\times$ & -- & $\times$ & $\times$ & $\times$ & $\times$ & $\times$ & $\times$ & $\times$ & \textit{$n_1$}:2, \textit{$n_2$}:6\\
\hline
\hline
\end{tabular}
\caption{\small The ansatz in (\ref{eq17}) can couple to the $D0^a$-brane here. The directions tangent and orthogonal to branes are denoted by -- and $\times$ respectively. The first two rows in the table are the original branes in the ABJM after taking near horizon limit. The six angles are $CP^3$ real coordinates, which are not required here. The numbers \textit{$n_1$} and \textit{$n_2$} show the number $\neq$ of \emph{relative transverse directions} of the added branes with the original ones as shown in the table. When there is one direction in the bulk of $AdS_4$, we have chosen it without any prefer to be $z$. Note also that in general the scalar function $f$ may depend on one, some or all directions of $AdS_4$ orthogonal to the brane world-volume.} \label{table1}
\end{table}

Now, we note that with taking $N/k=\lambda=1/(8\pi^2 {\grave{k}}^2)$ in (\ref{eq11}), we have a (anti) self-dual configuration $\widetilde{F}^{D2}=\mp \ast_4 F^{D0}$- where one may take the new symbol $\hat{F}_2^{\pm}$ for this (anti) self-dual 2-form. This is a valid approximation in the type IIA version of the ABJM when $\grave{k}$ is small and for completeness $\grave{k}\rightarrow0$. The good with this solution is the indices of $F_2$ and $F_4$ do not contract with those of the background fields. Further, the energy-momentum along $AdS_4$ vanishes. By taking $\mu,\nu,\ldots$ for the external space $AdS_4$ indices and $m,n,\ldots$ for the internal space $CP^3$ indices, $T_{\mu\nu}^{F_2}=0$ since $\hat{F}_2$ is (anti) self-dual. For $F_4$ we have as well
\begin{equation}\label{eq18}
  \begin{split}
   T_{\mu\nu}^{F_4} &=\frac{1}{2.4!}\big[4.3 F_{\mu\rho mn}F_{\nu}^{\rho mn}-\frac{1}{2}.6 g_{\mu\nu}F_{\sigma\rho mn}F^{\sigma\rho mn}\big]\\
   &=\frac{3}{2.4!} \big[4 F_{\mu\rho}F_{\nu}^{\rho}J_{mn}J^{mn}- g_{\mu\nu}F_{\sigma\rho}F^{\sigma\rho}J_{mn}J^{mn}\big]\\
   &=\frac{3}{4!} \big[2 F_{\mu\rho}F_{\nu}^{\rho}- \frac{1}{2}g_{\mu\nu}F_{\sigma\rho}F^{\sigma\rho}\big].J^2= \frac{J^2}{2!} T_{\mu\nu}^{F_2}=0
 \end{split}
\end{equation}
where the coefficients 3 and 6 in the first line are coming from the permutations of indices. Further, by using the complex coordinate on $CP^3$, for the energy-momentum tensor of $F_2$ along $CP^3$ we have
\begin{equation}\label{eq19}
T_{m\bar{n}}^{F_2}=-\frac{1}{8} g_{m\bar{n}} F_{\mu\nu}F^{\mu\nu}
\end{equation}
whereas, for $F_4$ it reads
\begin{equation}\label{eq20}
T_{m\bar{n}}^{F_4}=-\frac{3}{2.4!} \big[4 F_{\mu\nu}F^{\mu\nu}J_{m\bar{p}}J_{\bar{n}}^{\bar{p}}- g_{m\bar{n}}F_{\mu\nu}F^{\mu\nu} J^2\big]=+\frac{1}{8} g_{m\bar{n}} F_{\mu\nu}F^{\mu\nu}
\end{equation}
where the use is made of
\begin{equation}\label{eq21}
J_{m\bar{p}}J_{\bar{n}}^{\bar{p}}=g_{m\bar{n}}, \qquad J^2=2J_{m\bar{n}}J^{m\bar{n}}=6
\end{equation}
Therefore we have
\begin{equation}\label{eq22}
T_{m\bar{n}}^{F_2}+T_{m\bar{n}}^{F_4}=0
\end{equation}
We see that with the ansatz (\ref{eq8}) and case of (anti) self-dual configuration $\hat{F}_2^{\pm}=0$, the energy-momentum of the D0-D2 branes vanishes. This means the Einstein equation (\ref{eq7}) in the ABJM background doesn't change because of adding these new terms or branes.

Then we evaluate the corrected value of the action based on our solution (\ref{eq15a}). The relevant part of the original action (\ref{eq1}) now is
\begin{equation}\label{eq23}
  \grave{S}_{IIA}=-\frac{1}{4\kappa^2} \int \big[F_2 \wedge \ast F_2+ \widetilde{F}_4 \wedge \ast \widetilde{F}_4 \big]
\end{equation}
By using the Hodge-duals in (\ref{eq10b}) and (\ref{eq10c}), it reads
\begin{equation}\label{eq26}
  \grave{S}_{IIA}=\frac{1}{4\kappa^2}\frac{3R^9}{2k} Vol(AdS_4\times CP^3)+\frac{1}{(2\pi)^4} \frac{R^9\grave{k}^2}{48k^3} \int {F}^{D0} \wedge \ast_4 {F}^{D0}
\end{equation}
where $Vol(\ldots)$'s are for the unit volume-elements and use is made of $\acute{\alpha}=1\to\kappa^2=\frac{1}{2}(2\pi)^7$ and $\int_{CP^3} J^3=(2\pi)^3$. The first term is the original one of the ABJM we call it $S_0$, while the second term is the correction induced by the solution (\ref{eq16}). For the latter, we write
\begin{equation}\label{eq27}
  S_{modi.}^{D2}=\frac{1}{(2\pi)^4}\frac{R^3}{12k} \int \widetilde{F}^{D2} \wedge \ast_4 \widetilde{F}^{D2}=\frac{1}{4} S_{modi.}^{D0}=\frac{1}{(2\pi)^4}\bigg(\frac{R^9\grave{k}^2}{16.12k^3}\bigg) \int_{AdS_4} (\vec{\nabla}f)^2 d^3r du
\end{equation}
From this, to have a finite correction, the objects must locate in some parallel planes orthogonal to the $u$ direction, say $u_{\to\infty}=\Lambda$. Then we may write
\begin{equation}\label{eq28}
 \int_{AdS_4} (\vec{\nabla}f)^2 d^3r du =\Lambda \int_{R^3} (\partial_k f) (\partial^k f)\approx-\int_{R^3} f \partial^2 f=-\int_{S^2} f \partial_k f d\Sigma^k
\end{equation}
We now proceed same as in \cite{Gibbons} to evaluate the value of the action. Using the clear solution (\ref{eq15a}) and noting the contribution from $r=0$ is vanished, we have just contribution of $r=\infty$, which is $4\pi c$ for the latter equation by taking $c_1=1$ as well. Therefore we can write
\begin{equation}\label{eq29}
 S_{m.inst} =4\pi Q_0,\qquad Q_0=\frac{c\Lambda \grave{k}^2}{(2\pi)^7}
\end{equation}
where $Vol(S^2)=4\pi$ uses for the unit 2-sphere and the full volume of Vol($CP^3(\tilde{R})$)$=\pi^3 R^9/6k^3$ is factored out as a normalization coefficient. This value of action should not be confusing, compared with the $1/g_s$-dependent actions for D-branes, since the monopole here is of the Dirac type\footnote{One may phrase this another way. In fact, based on $f$ in (\ref{eq15a}), one can write $\int_{R^3} (\partial f)^2 \sim \frac{Q_0}{\epsilon}$, where $\epsilon$ is the radius of an infinitesimal sphere surrounding the monopole instanton-like object.}.

Now we evaluate the electric charge associated with the D0-brane through
\begin{equation}\label{eq30}
 Q^{D0}=\frac{1}{\sqrt2 \kappa} \oint \ast F_2^{D0}=\frac{\grave{k}R^9}{48\sqrt{2\pi}k^3}\oint \ast_4 F^{D0}=-\frac{c \grave{k}}{8(2\pi)^{5/2}}
\end{equation}
where $F_2^{D0}$ implies the second term of $F_2$ in (\ref{eq8}) and that we have used the explicit form of the field in (\ref{eq16}) to perform integration in the second line with integrating out the $CP^3$ volume. Further, from above, with $c=1$ for one monopole or even more, it is obvious the charge is small compared with the ABJM background fluxes. This means neglecting back-reactions as we have shown it clearly of course. In other words, from the correction (\ref{eq29}) to the ABJM action and the latter charge we obviously see the contributions are small compared with the original background. That is because almost $\grave{k}\rightarrow0$, which is a valid statement especially for (anti) self-dual solution outlined above. As aside note also that if we take $\Lambda=1$ for convenience, the action as well reads
\begin{equation}\label{eq31}
 S_{m.inst}=\frac{64|Q^{D0}|^2}{\pi c}
\end{equation}
Last and to connect the discussion to the next section, we note of the relevant decomposing of the gauge group as surveyed first in \cite{NilssonPope}\footnote{ Look also at \cite{Bianchi2} including references therein for a rederivation of the spectrum with respect to the ABJM.}. In fact, by considering $S^7$ as a $U(1)$ fibration on $CP^3$, the \textbf{28} representation of $SO(8)$ for the gauge fields decomposes as $\textbf{28}\rightarrow \textbf{1}_0+\textbf{6}_2+\textbf{6}_{-2}+\textbf{15}_0 $ under $SO(8) \rightarrow SU(4) \times U(1)$. Therefore, the only remaining gauge bosons are in $\textbf{1}_0+\textbf{15}_0$ and neutral under $U(1)$. The singlet is one needed for us on its exact form we concentrate in what follows.

\section{Field Theory Side and Correspondence}
For a $p$-form with the mass squared $m^2$ in $AdS_{d+1}$, having the radius curvature $L$, relation between the mass $m$ and the scaling-dimension $\Delta$ is as $m^2L^2=(\Delta-p)(\Delta+p-d)$. Thus, for the massless gauge field $A^{D0}$ in $AdS_4$, corresponding boundary theory involves the operators of the conformal-dimension $\Delta_{\pm}=1,2$. For supergravity multiplets of the lowest mass, only the upper branch $\Delta_+=2$, which is the normalizable mode, is suitable. Therefore, we should search for this invariant operator- indeed singlet- under $SU(4)_R \times U(1)_b$ since the gravity ansatz (\ref{eq16}) has this property. We have already had some discussions \cite{I.N} on the operators of the conformal dimensions $\Delta_{\pm}=1,2$ when we were dealing with a massive scalar in the bulk. It was a conformally coupled scalar. There we forced to turn on a Fermi field next to a gauge filed of $U(1) \times U(1)$ to match both sides of duality. We now see the aim is achieved here similarly by turning on scalars next to the gauge fields of the boundary theory.

The action of the ABJM model is given in \cite{ABJM} and also \cite{Klebanov}. The field content of the ABJM action consists of two gauge fields $A_i$ and $\hat{A}_i$ taking value in the Lie algebra of the gauge group $U(N) \times U(N)$, bi-fundamental bosonic fields $Y^A$ ($A=1,..,4$) and spinor fields $\psi_A$ in the \textbf{${\textbf{4}}$} of representation $SU(4)$ global R-symmetry. The explicit $SU(4)_R$ invariant action reads
\begin{equation}\label{eq32}
 \begin{split}
   S_{ABJM} =\int d^3r \bigg\{\frac{ik}{4\pi} \varepsilon^{ijk} & tr \bigg(A_i \partial_j A_k+\frac{2i}{3} A_i A_j A_k-\hat{A}_i \partial_j \hat{A}_k-\frac{2i}{3} \hat{A}_i \hat{A}_j \hat{A}_k \bigg)\\
   & -tr\big(D_k Y_A^\dagger D^k Y^A \big)-tr\big(\psi^{A\dagger} i \gamma^k D_k \psi_A \big)-V_{bos}-V_{ferm} \bigg\}
 \end{split}
\end{equation}
where
\begin{equation}\label{eq34}
   \begin{split}
   V_{bos} =-\frac{4\pi^2}{3k^2} tr&\big(Y^A Y_A^\dagger Y^B Y_B^\dagger Y^C Y_C^\dagger + Y_A^\dagger Y^A Y_B^\dagger Y^B Y_C^\dagger Y^C + 4 Y^A Y_B^\dagger Y^C Y_A^\dagger Y^B Y_C^\dagger \\
    & - 6 Y^A Y_B^\dagger Y^B Y_A^\dagger Y^C Y_C^\dagger\big)
  \end{split}
\end{equation}
 \begin{equation}\label{eq35}
   \begin{split}
   V_{ferm} =-\frac{2\pi i}{k} tr&\big(Y_A^\dagger Y^A \psi^{B\dagger}\psi_B - Y^A Y_A^\dagger \psi_B \psi^{B\dagger} + 2 Y^A Y_B^\dagger \psi_A \psi^{B\dagger} -2 Y_A^\dagger Y^B \psi^{A\dagger}\psi_B \\
   & +\varepsilon^{ABCD} Y_A^\dagger \psi_B Y_C^\dagger \psi_D- \varepsilon_{ABCD} Y^A \psi^{B\dagger} Y^C \psi^{D\dagger}\big)
   \end{split}
 \end{equation}
Note the $i$ factor in front of the Chern-Simons term due to being in the Euclidean space.

Setting fermions to zero, equation of motion for $Y_A^\dagger$ reads
\begin{equation}\label{eq35}
   \begin{split}
    D_k D^k Y^A &=-\frac{4\pi^2}{k^2} \big\{(Y^C Y_C^{\dagger}) (Y^B Y_B^{\dagger})Y^A+Y^A (Y_B^{\dagger}Y^B) (Y_C^{\dagger}Y^C)+4 Y^C Y_B^{\dagger}Y^A Y_C^{\dagger} Y^B \\
    &-6 Y^A Y_B^{\dagger} (Y^C Y_C^{\dagger})Y^B-2 (Y^C Y_C^{\dagger})Y^A (Y_B^{\dagger}Y^B)-2 Y^C (Y_B^{\dagger}Y^B) Y_C^{\dagger}Y^A \big\}
   \end{split}
\end{equation}
and its dagger for $Y_A$. For the gauge fields $A_i, \hat{A}_i$ the equations, known as the Gauss's law constraints, are
\begin{equation}\label{eq36}
    \begin{split}
    & \frac{ik}{4\pi} \varepsilon^{kij} F_{ij}= i\big[Y^A (D^k Y_A^{\dagger})-(D^k Y^A)Y_A^{\dagger} \big], \\
    & \frac{ik}{4\pi} \varepsilon^{kij} \hat{F}_{ij}= i\big[(D^k Y_A^{\dagger})Y^A- Y_A^{\dagger} (D^k Y^A) \big]
    \end{split}
\end{equation}
where
\begin{equation}\label{eq33}
   \begin{split}
   & D_k Y^A =\partial_k Y^A+iY^A (A_k -\hat{A}_k), \\
   & F_{ij}=\partial_i A_j-\partial_j A_i+i \big[A_i, A_j \big]
   \end{split}
 \end{equation}
and Noether current for U(1) gauge transformations, called baryonic symmetry, is
\begin{equation}\label{eq37}
    J_b^k=-tr\big[Y^A (D^k Y_A^{\dagger})-Y_A^{\dagger} (D^k Y^A) \big], \\
\end{equation}
Reviewing the needed materials, we are now ready to find the equivalent solution and confirm the correspondence. As our gravity solution preserves some supersymmetry, we should search for the matching BPS solution here. If we turn on just one of the four scalar fields say $Y^1$, from $D_k Y^A $ in (\ref{eq33}) or its dagger, we have
\begin{equation}\label{eq38}
    DY^1=0 \quad or \quad DY_1^\dagger=0
\end{equation}
This is condition for half-BPS configurations equivalent to the BPS equation coming from the $\psi$ transformations as discussed for instance in \cite{Terashima, Gomiz} and also \cite{Lin}. If we take $Y_1^{\dagger}$ to be the complex conjugate of $Y^1$, solution is trivial. But we note that, in the Euclidean space, they can be treated independently as our correspondence confirms it too. Now, by taking $DY^1=0$, without any prefer, and the Gauss constraints (\ref{eq36}), it is easy to see the equation of motion reads
\begin{equation}\label{eq39}
   D_k D^k Y_1^\dagger=0
\end{equation}
that is same equation coming from (\ref{eq35}) in that with only one scalar field turned on, the scalar potential vanishes. By making the following ansatz
\begin{equation}\label{eq40}
   Y^1=c_3\textbf{1}, \qquad  Y_1^\dagger=h(r)\textbf{1}
\end{equation}
where $c_3$ is a constant its value will fix and \textbf{1} is the unit matrix. Neglecting the gauge fields for now, solution for this real $h(r)$ is same as that for $f(r)$ in (\ref{eq15a}). But this is not full story yet. What is the exact form of the agreeing $SU(4)$ singlet operator sourced by the bulk gauge field $A^{D0}$. Further, the counterpart to the D0-brane charge in (\ref{eq30}) is still missing here. We see that to match both side solutions, the boundary dimension-2 singlet operator can be constructed from the $U(N) \times U(N)$ gauged fields. In fact, considering two $U(1)$'s with discarding other $SU(N)$ gauge fields, as done in \cite{I.N} as well, take us on the right way and make everything consistent.

The scalars and fermions, in general, couple to a special combination of the gauge fields. Therefore, introducing the symbol $A_i^\pm\equiv (A_i\pm\hat{A}_i)$ is convenient. The fundamental fields $Y^A$, which are natural under the diagonal $U(1)$, couple to $A_i^+$ whereas the orthogonal combination $A_i^-$ acts as the baryonic symmetry. Thus, from (\ref{eq36}) we can write
\begin{equation}\label{eq41}
    \begin{split}
     \frac{k}{4\pi} \varepsilon^{kij} &F_{ij}^+=\big(Y^A (D^k Y_A^{\dagger})-(D^k Y^A)Y_A^{\dagger} \big), \\
     &F_{ij}^-=0
    \end{split}
\end{equation}
Further we set $A_i^-=0$. On the other hand, we can write
\begin{equation}\label{eq42}
    F_{ij}^+=\partial_i A_j^+-\partial_j A_i^+=\varepsilon_{ijk}B^k=-\varepsilon_{ijk}\partial^k \phi_m
\end{equation}
where $\phi_m$ is the scalar magnetic potential. If we identify $\phi_m$ with $f$, the resulting magnetic field $\vec{B}$ is nothing but the familiar magnetic field of a point charge located at $r=0$. Plugging the latter expression with (\ref{eq40}) in (\ref{eq41}) and noting that $f(r)=h(r)=\phi_m$, we arrive in
\begin{equation}\label{eq43}
    c_3=-\frac{k}{2\pi}
\end{equation}
From (\ref{eq37}), with (\ref{eq40}) and the latter relation, the associated current and charge for the $U(1)$ gauge transformation read
\begin{equation}\label{eq43a}
    J_k=-\frac{k}{2\pi} \partial_k h(r), \quad Q_b=\int J_k d\Sigma^k=2kc
\end{equation}

By considering the conformal dimension-2 operator as $\mathcal{O}_2\sim F^+$, the bulk field $A^{D0}$ couples through $W\sim\int_{R^3} A^{D0} \wedge F^+\sim S_{WZ}^{D2}$ as we explain now. Indeed for the $U(1) \times U(1)$ theory, which is the case with one D2-brane, the Chern-Simons term reads
\begin{equation}\label{eq44}
    S_{CS}=\frac{ik}{4\pi}\int d^3r \,\varepsilon^{ijk} A_i^- F_{jk}^+
\end{equation}
By defining the theory on $R^3$ and projecting it to $R \times S^2$, there are sections having $\int_{S^2}F^+=4\pi c$. Now our deformation as $S\rightarrow S+W$ with $W=-S_{modi.}^{D0}$, because of turning on the bulk gauge field $A^{D0}$, equal to a gauge transformation as
\begin{equation}\label{eq45}
  A_i^-\rightarrow A_i^-+\partial_i \tilde{f} \equiv A_i^-+\beta A_i^{D0}
\end{equation}
where $\tilde{f}=\alpha f$ with $\alpha$ (and $\beta=2\alpha$) as some constant its value will fix\footnote{Note  that under this $U(1)_b$ transformation, $Y^A\rightarrow e^{\tilde{f}}Y^A$ - and similar for $\psi_A$, which is of course settled to zero here.}. Therefore, making use of (\ref{eq42}), the boundary term induced by this transformation is
\begin{equation}\label{eq46}
    W_{m.inst}=-\frac{ik\alpha}{2\pi}\int_{S^3} d^3r \, (\partial_k f)(\partial^k f)
\end{equation}
The integral measure is same as that in (\ref{eq28}) with $\Lambda=1$. Comparing both sides with noting that $W_{m.inst}=-S_{m.inst}$, result in
\begin{equation}\label{eq47}
    \alpha=-\frac{i\grave{k}^2}{k(2\pi)^6}
\end{equation}

Indeed, from the viewpoint of the D2-brane world-volume action $S_{WZ}^{D2}$, the external gauge field $A^{D0}$ couples to the background D2-brane in  above standard way. This D0-instanton can be considered as an intersection in the context of $Dp-D(p+2)$ bound sates. It seems that we have a similar interpretation as in \cite{Hosomichi}, where the founded monopole-instanton solution was a D0-brane mediating two D2-branes. The D0-brane here, as a source for $A^{D0}$, may have similar role. Now, this Euclideanized D0-brane having monopole instanton-like character, may interpolate between one background Euclideanized D2-brane and an added Euclideanized D2-brane whose associated field is given by $F_4$ in (\ref{eq8}). Even more related interpretation is a D0-brane smeared in the original D2-branes. Lifting to M-theory, this D0-brane can be interpreted as existing some charges of the KK-modes along the $S^1$ direction associated with the fiber coordinate. The latter interpretation may also be considered as a M0-brane discussed in \cite{Terashima2} as well.

\section{Concluding Remarks}
Having the solution in the last section, here we follow the brief discussion in the introduction about electric-magnetic duality. This study is around the previous works \cite{Witten2}, \cite{Yee} and \cite{deHaro} on the S-duality of the Abelian gauge fields in AdS since we have been facing a similar case. The procedure outlined above is plainly the Dirichlet condition. Because the boundary value of the magnetic field fixes and $A_i^-$ is a source for a symmetry current in CFT. More clearly, the gauge field $A_i^-$ couples to a U(1) current $J^i$ and therefore the new field theory Lagrangian $\tilde{L}$ includes a new term $A_i^- J^i$ next to scalars in the original Lagrangian $L$, i.e., $\tilde{L}=L+ A_i^-J^i$. In other words, from $F_{ij}^+$ in (\ref{eq42}), $J^k$ in (\ref{eq43a}) and then (\ref{eq44}), we can write
\begin{equation}\label{eq48}
   F_{ij}^+=-\frac{2\pi}{k}\varepsilon_{ijk}J^k \rightarrow S_{CS}=-i \int A_k J^k
\end{equation}
This is Dirichlet condition and usual CFT in the language of \cite{deHaro}. On the other hand, similar to the scalars, as approaching to the boundary at $u=0$, we can write for a gauge field $A$
\begin{equation}\label{eq49}
   A(u,\vec{u})=A^-u + F^+ u^2
\end{equation}
the usual CFT is one that couples to the source $A_i^{D0}\sim A_i^-$ and an operator of the conformal dimension $\Delta_+=2$ on which we have been concentrating above. We indeed have $S[A^{D0}]=-W[A_i^-]$ and $\langle \mathcal{O}_2 \rangle_{A_i^-}=F^+$. The two-point function of this is also evaluated in the mentioned three references.

Now, to do the S-operation, one should first gauge the U(1) global symmetry, then promote the gauge field $A_i^-$ to a dynamic one and third couple it to an external gauge field ($A^{D0}$ here) through a Chern-Simons coupling. After making this Legendre transformation we meet a dual CFT. In \cite{deHaro}, the interplay among S-duality, Legendre transformation and RG flow is discussed as well. It is notable that under this Renormalization Group flow to InfraRed, the IR theory is described by the dimension-2 current $J$, which is S-dual to the UV current $\tilde{J}$. The latter current comes because $A_i^-$ cannot be an operator by itself. Then one can make the dual current $\tilde{J}=\ast_3 dA_i^-=\vec{B}$, which has the dimension-2 and satisfies the unitary bound $\Delta\geq2$. The gauge field $A_i^+$ is now playing the role of source.

Both theories are related by a Legendre transformation through an AB-type Chern-Simons term $\sim \int A^- \wedge dA^+$. Indeed our study here may be seen as the particle-vortex duality in the upper row of the Figure 1 of \cite{deHaro}. Meanwhile, we note the vertical case there is for a Chern-Simons coupling between a background field ($A_i^+$ here) and a new dynamic gauge field ($A^{D0}$ here). Therefore, our instanton-like object may also well-adjusted to the left column of that figure. There are also some discussions on the self-dual boundary conditions replying to the bulk self-dual solutions, which is a special case of our solution.

Finally and to summary, starting with an ansatz for the form fields in the type IIA gravity side of the ABJM model, we arrived in a localized object in the bulk of $AdS_4$. We referred the solution as a monopole-instanton like object (indeed a D0-instanton). We saw that, in a special limit, our solution in the bulk could be (anti) self-dual. In this limit it has the property of being an exact solution without any back-reaction on the geometry as a topological object must have it. Then we evaluated the relevant part of the supergravity action and charge of object based on our exact solution. The latter led us to an understanding that our monopole is of the Dirac type. Afterwards, knowing that we have a singlet U(1) gauge field in the bulk coming from a consistent Kaluza-Klein dimensional reduction on the associated space, we searched for the dual BPS solution and operator on the boundary. In doing so, we turned on just one scalar field on the boundary next to the $U(1) \times U(1)$ part of the gauge group. Then, by making use of symmetries, the corresponding operators came out and both side solutions matched as well. Last, we represented some comments on the electric-magnetic or S-duality of our case based on the earlier works confirming our procedure too.

\section{Acknowledgements}
I would like to thank A. Imaanpur for his collaboration in an early stage of this study.

\end{document}